\documentstyle[12pt]{article}
\begin{document}
\begin{titlepage}
\begin{flushright}
Z\"urich University ZU-TH 15/94\\
Pavia University FNT-T 94/22\\
\end{flushright}
\vfill
\begin{center}
{\large\bf A SCENARIO FOR A BARYONIC DARK HALO}\\
\vfill
{\bf F. De Paolis$^{1,2,\#}$, G. Ingrosso$^{1,2,\#}$,
Ph.~Jetzer$^{3,}$* and M. Roncadelli$^4$}\\
\vskip 1.0cm
$^1$Dipartimento di Fisica, Universit\'a degli Studi di Lecce,
Via Arnesano CP 193,\\
 I-73100 Lecce, Italy,\\
$^2$INFN, Sezione di Lecce, Via Arnesano CP 193, I-73100 Lecce,
Italy,\\
$^3$Institute of Theoretical Physics, University of Z\"urich,
Winterthurerstrasse 190,\\
CH-8057 Z\"urich, Switzerland,\\
$^4$INFN, Sezione di Pavia, Via Bassi 6, I-27100 Pavia, Italy.
\end{center}
\vfill
\begin{center}
Abstract
\end{center}
\begin{quote}
The recent observations of microlensing events in the LMC by the MACHO
and EROS collaborations suggest that an important fraction of the
galactic
halo is in the form of Massive Halo Objects (MHO) of about $0.1
M_{\odot}$.
Here, we argue that the galactic halo is mainly baryonic and that
besides MHO also H$_2$ molecular clouds may significantly contribute to
it.

We propose a scenario in which dark clusters of MHO and/or
H$_2$ molecular clouds
form in the halo at galactocentric distances larger than $\sim 10-20$
kpc,
since there
we expect less collisions among  proto globular cluster clouds and
a smaller UV radiation flux.

Cosmic ray protons may induce a significant $\gamma$-ray flux
in $H_2$ molecular clouds. Our calculation gives an upper bound
to this flux which is below present detectability.\\

{\bf Key words:} dark matter - gravitational lensing - galaxies:
evolution
- ISM: clouds - ISM: molecules - gamma rays: theory
\end{quote}
\vfill
\begin{flushleft}
\vfill
\begin{center}
June 1994
\end{center}
\vskip 0.5cm
$^*$ Supported by the Swiss National Science Foundation.\\
$^{\#}$ Partially supported by Agenzia Spaziale Italiana.
\end{flushleft}
\end{titlepage}
\newpage
\section{Introduction}
One of the most important problems in astrophysics concerns
the nature of the dark
matter in galactic halos, as suggested by the observed
flat rotation curves.
Present limits coming from primordial
nucleosynthesis allow a halo made of ordinary baryonic matter
that should be in the
form of Massive Halo Objects (MHO) with masses in the range
$10^{-7} < M/M_{\odot} < 10^{-1}$ (De R\'ujula et al. \cite{Deruj}).
They
can be detected
using the gravitational lens effect (Paczy\'nski \cite{Pa1}).

Recently, the French collaboration EROS (Aubourg et al. \cite{Aubourg})
and the
American-Australian collaboration MACHO (Alcock et al. \cite{Alcock})
reported
the possible detection of altogether six microlensing events,
discovered by
monitoring over several years millions of stars in the Large Magellanic
Cloud, whereas the Polish-American collaboration OGLE (Udalski et al.
\cite{Uda}) and also the MACHO collaboration have found thirteen
microlensing
events by monitoring stars located in the galactic bulge.
Taking these results at face value, an average mass
$\sim 0.1 M_{\odot}$ for the MHO in the halo has been derived
(Jetzer 1994).
At present, it is very difficult to estimate the fraction of dark
matter in
the halo which could be in the form of MHO, due to the very limited
number of
observations at disposal.
Nevertheless, we feel it hard to conceive that a
formation mechanism exists which transforms with 100\% efficiency
hydrogen and
helium gas into MHO. Thus, we expect that also H$_2$ molecular clouds
should
be present, since clouds of neutral H gas would be detected through the
21-cm
radioband frequency. Of course, it is still possible that a fraction of
the
dark matter is non baryonic.

Pfenniger et al. \cite{Pfe1} suggested that H$_2$ molecular clouds
could
make up the dark matter in the disk of the Galaxy and Lequeux et al.
\cite{Lequeux}  recently
reported the detection of faint CO lines from gas at a distance of
about
12 kpc from the galactic center. This observation suggests that there
is a
significant amount of H$_2$ molecular gas in the outer regions of the
Galaxy.
Here, we argue that the halo of our galaxy is mainly baryonic
and made of dark clusters
of MHO and/or H$_2$ molecular clouds.

\section{Dark cluster formation}
We suggest that dark clusters form at large galactocentric distances in
a more
quiet phase of the collapsing proto Galaxy (PG).
The proposed scenario relies on the theory for the origin of globular
clusters
advocated by Fall and Rees \cite{Fall}
 and on the suggestion of Palla, Salpeter \&
Stahler \cite{Palla1} that the lower bound on the fragment masses in a
collapsing, metal poor cloud can be as low as $10^{-2} M_{\odot}$.

Fall and Rees \cite{Fall} argued that overdense regions in the PG cool
more
rapidly than average
and then a two phase medium forms with cool proto globular cluster
(PGC) clouds in pressure equilibrium with the external
hot diffuse gas. The PGC clouds, which have temperature $T_c \sim
10^4~K$,
are gravitationally unstable when their mass exceeds the
corresponding Jeans mass.

The main coolants below $10^4~K$ are molecular hydrogen and any heavy
element
produced by a first generation of stars. Molecular hydrogen, however,
would be
dissociated by various sources of UV radiation such as  an AGN and/or a
population of massive young stars in the center of the PG.

In a metal poor, dust free protogalactic gas
the molecular hydrogen abundance $f_{H_{2}}=n_{H_2}/n_H$ is determined
by
considering a set of atomic and molecular processes
which involve the creation of intermediate $H^-$ and $H_2^+$
(see, e.g., Shapiro and Kang \cite{Sha}).
With the knowledge of $f_{H_2}$ one can compute the cooling time
$\tau_{cool}= 3\rho_c k_B T_c/[2\mu_c (\Lambda_c -\Gamma )]$,
where $\rho_c$ is the density of the PGC cloud, $\mu_c$
the mean molecular weight, while
$\Gamma$ and $\Lambda_c$ are
the heating and cooling rates, respectively.
The subsequent evolution of the PGC cloud depends on the ratio between
$\tau_{cool}$ and the gravitational infall time
$\tau_{ff}$.
The parameters that mainly discriminate
between the different situations are $\Gamma$ and $\Lambda_c$.

In the inner halo, because of the presence of
an AGN and/or a first population of massive stars
at the center of the PG,
$H_2$ formation and cooling are heavily suppressed or delayed for a
wide
range of both external UV fluxes and/or PGC cloud densities
(Kang et al. \cite{K1}).
For this case, in which $\tau_{cool}\geq\tau_{ff}$,
the PGC clouds remain at $T_c\sim 10^4~K$ for a long time. This results
in an {\em imprinting} of a characteristic mass of
$M_c \sim 10^6~M_{\odot}$.
After enough $H_2$ has formed ($f_{H_{2}}\sim 10^{-3}$),
the temperature suddenly drops well below $10^4~K$,
because now $\tau_{cool}\ll \tau_{ff}$.
The subsequent evolution of the PGC clouds goes on with a rapid grow of
the
small scale
perturbations that leads directly (in one step)
to the formation of stars within a narrow mass range (Murray \& Lin
\cite{Murray1}).
This scenario would explain
the formation of stellar globular clusters.

In the outermost part of the halo, where the incoming UV radiation flux
diminishes due to the distance (so that $\tau_{cool}\leq\tau_{ff}$),
the
PGC clouds cool more gradually below $10^4~K$. Then, cooling and
collapse
occur simultaneously and the evolution of the PGC clouds proceeds
according
to the scenario proposed by Palla et al. \cite{Palla1}, leading to a
subsequent fragmentation into smaller clouds that remain optically thin
until
the minimum value of the Jeans mass ($\leq 0.1 M_{\odot}$) is attained.
In
this case, because the PGC clouds do not remain at a fixed temperature,
there
should be no imprinting of a characteristic mass. In fact, when a cloud
is in
a quiet ambient as at the edge of the PG, the collapse proceeds with
a monotonically decrease of the Jeans mass and a subsequent
fragmentation
into clouds with lower and lower masses. This process stops when the
fragments
become optically thick to their own line emission.
This happens because in a metal poor gas and over a wide range of
initial
conditions, starting from densities $n_H > 10^{8}~cm^{-3}$,
three body reactions
($H + H + H  \rightarrow H_2 + H$ and
 $H + H + H_2\rightarrow 2H_2   $) are so efficient that
the atomic hydrogen is rapidly converted into molecular form.
As a consequence
of the increased cooling efficiency the fragmentation processes
go on until the Jeans mass reaches its minimum value.
The fragmentation is favoured if
the gas has been metal enriched due to a primordial
star population (Palla and Stahler \cite{Palla2}).

The result of the above picture would be the formation in the galactic
halo
of dark clusters made of MHO with mass $\sim 0.1~M_{\odot}$ or less.
However, we don't expect the fragmentation process to be able
to convert the
whole gas mass contained in a PGC cloud into MHO.
For instance, standard stellar formation mechanisms lead to upper
limits of at most $40\%$ for the conversion efficiency (Scalo
\cite{Scalo}).
Thus, we expect the remaining gas to form self-gravitating $H_{2}$
molecular
clouds, since in the absence of strong stellar winds the surviving gas
remains
gravitationally bound in the dark clusters.
The possibility that the gas is diffuse in the dark clusters is
excluded due to its high virial temperature ($\sim 10^4$ K)
that would make the gas observable at 21 cm.
In addition, the gas cannot have diffused in the whole galactic halo
because it would have been heated by the gravitational
field to a virial temperature $\sim 10^7~K$ and, therefore, would be
observable in the X-ray band (for which stringent upper limits are
available).

As noted previously, a sufficiently high UV flux is needed to form
stellar
globular clusters. Hence, they can mainly form up to a certain
galactocentric
 distance
beyond which the evolution of the PGC clouds gives rise to the
formation of dark clusters of MHO and/or $H_2$ molecular clouds.
For a UV flux due to a central AGN this critical distance
is $R_{crit}^{AGN} \simeq (L_{AGN}/2 \times 10^{42} erg~s^{-1})^{1/2}$
kpc.
If instead the UV flux is produced by early massive stars mainly
located at
the center of the PG, we get
$R_{crit}^{\star}\simeq 10^{-3}~L_{tot}/[L_{\star}~n_{\star}(0)]$ kpc,
where $L_{\star}$ is the bolometric luminosity of a single B0 V star,
$n_{\star}(0)$ the central stellar density and $L_{tot}$ the total
stellar
luminosity.
By using numerical values in tables 1 and 2 of Kang et al. (\cite{K1})
we estimate $R_{crit}^{AGN} \sim 20$ kpc
and $R_{crit}^{\star} \sim~ 10$ kpc.

A further question which arises is whether dark clusters are stable
within the lifetime $t_g$ of the Galaxy. Standard calculations
for the evaporation time require a constituent mass $m \leq 10~
M_{\odot}$
in order to have $t_{evap} > t_g$ (Carr and Lacey \cite{Carr}).
Another mechanism which could destroy dark clusters are collisions
among
themselves. Dark clusters will be disrupted over the lifetime $t_g$ if
they
are located inside a radius $R_{dis} \sim 6$ kpc (Carr and Lacey
\cite{Carr}).
Therefore, we conclude that dark clusters made of MHO and/or $H_2$
molecular
clouds can still be present today in the galactic halo, provided
they are located
at distances larger than $R_{dis}$.

We mention that the possibility to infer from microlensing observations
whether MHO are clustered or not has recently been considered by Maoz
\cite{Maoz}.
In the former case, in fact, the degeneracy in the spatial and velocity
distributions of MHO
would result in a strong autocorrelation in the sky position of
microlensing
events on an angular scale of 20 arcsec, along with a correlation in
the
event duration.

\section{Bounds on the
$\gamma$-ray flux produced by halo molecular clouds}
Here we estimate the $\gamma$-ray flux produced in
H$_2$ molecular clouds located in the halo through the interaction
with high-energy cosmic-ray
protons.
Cosmic rays scatter on protons in the H$_2$ molecules producing
$\pi^0$'s,
which subsequently
decay into $\gamma$'s.
Outside the clouds we
expect negligible $\gamma$-ray photon ($\geq$ 100 MeV)
absorption in the interstellar medium, since
for the typical interstellar medium
density $\sim 1$ atom $cm^{-3}$ the a mean free path is about
20 Mpc.

An essential ingredient is the knowledge of the cosmic ray
flux in the halo. Unfortunately, this quantity is unknown
and the only information comes from theoretical estimates (see, e.g.,
Breitschwerdt et al. \cite{Breitschwerdt}).
More precisely, from the mass-loss rate of a
typical galaxy, one can infer the total cosmic ray flux in the halo,
which turns out to be
$F\simeq 10^{41}~ A_{gal}^{-1} ~erg~cm^{-2}~ s^{-1}$, where $A_{gal}$
is the
surface area of the galactic disk. Taking $R_{gal} \simeq$ 10 kpc, we
get
$F \simeq 1.1\times 10^{-4}~ erg~cm^{-2}~s^{-1}$.
However, this  information
is not sufficient to carry out our calculations, since we need the
energy
distribution of the
cosmic rays. Because this is not known, we assume the same energy
dependence
as measured on Earth. We then scale the overall
density in such a way that the integrated energy flux agrees with the
above
value. Moreover, we assume that the cosmic ray density scales as
$1/R^2$ for a large galactocentric distance $R$.
The measured primary cosmic ray flux on the Earth is
$\Phi^{\oplus}_{CR}(E)
\simeq 2~(E/GeV)^{-2.7}~particles~cm^{-2}~s^{-1}~sr^{-1}$
(see, e.g., Lang \cite{Lang}).
Therefore, the corresponding integrated energy flux
(we take the integration range to be $1~ GeV \leq E \leq 10^6~ GeV$) is
$F^{\oplus} \simeq 5.7\times 10^{-2}~ erg~cm^{-2}~s^{-1}$.
Comparison between the values of $F$ and $F^{\oplus}$
entails $\Phi_{CR}(E) \simeq 1.9\times 10^{-3}~\Phi_{CR}^{\oplus}$,
so that
- in accordance with our previous assumption - we get
\begin{equation}
\Phi_{CR}(E, R) \simeq \Phi_{CR}(E)~~\frac{a^2+R_{GC}^2}{a^2+R^2}~,
\label{eqno:45}
\end{equation}
where $a\sim 5$ kpc is the halo core radius and $R_{GC}\sim 8.5$ kpc is
our
distance from the galactic center. We notice that
our result exhibits a very mild $a$-dependence.

Let us now turn our attention to the evaluation of the $\gamma$-ray
flux
produced through the reactions $pp \rightarrow \pi^0 \rightarrow \gamma
\gamma$. Accordingly, the source function
$q_{\gamma}(r)$ giving the photon number density at distance
$r$ from the Earth is
\begin{equation}
\begin{array}{ll}
q_{\gamma}(r)&= \displaystyle{
\frac
{4\pi}
{m_p}
} ~ \rho_{H_2}(r) ~
\sum_{n} \\ \\ \hspace*{-40pt}& \int dE_p~ dE_{\pi}  \Phi_{CR}
(E_p,R(r)) \displaystyle{
\frac
{d \sigma^n_{p \rightarrow \pi} (E_{\pi})}
{dE_{\pi}}
}
n_{\gamma}(E_p)~, \label{eqno:46}
\end{array}
\end{equation}
where $\sigma^n_{p \rightarrow \pi}(E_\pi)$ is the cross section for
the
reaction $pp \rightarrow n \pi^0$ ($n$ is the $\pi^0$ multiplicity),
$n_{\gamma}(E_p) \simeq 2 + 1.03~ n_{ch}(E_p)$ is the photon
multiplicity
(Forti et al. \cite{Forti}),
while $R(r)$ is the galactocentric distance as a function of $r$.
Next, we re-express $q_{\gamma}(r)$ in terms of the inelastic pion
production
cross section. Since
\begin{equation}
\sigma_{in}(p_{lab})<n_{\gamma}(E_p)> = \sum_n \int dE_{\pi}
\frac{d\sigma^n_{p \rightarrow \pi}(E_{\pi})}{dE_{\pi}}
n_{\gamma}(E_p),
\label{eqno:48}
\end{equation}
Eq. (\ref{eqno:46}) becomes
\begin{equation}
\begin{array}{ll}
q_{\gamma}(r)&= \displaystyle{
\frac{4\pi}{m_p}}~ \rho_{H_2}(r) \\ \\ &
\int dE_p~ \Phi_{CR}(E_p,R(r))
{}~\sigma_{in}(p_{lab}) <n_{\gamma}(E_p)>~. \label{eqno:49}
\end{array}
\end{equation}
Actually, the cosmic ray protons in the halo
which originate from the galactic disk are mainly directed outwards.
This fact
implies that also the induced photons will leave the Galaxy.
However, the presence of magnetic fields in the halo might give rise
to a temporary confinement of the cosmic ray protons similarly to what
happens in the disk.
In addition, there could also be sources of
cosmic ray protons located in the halo itself, as for instance
isolated or binary pulsars in globular clusters.
Unfortunately, we are unable to give a quantitative estimate of the
above effects, so that we take them into account by introducing an
efficiency factor $\epsilon$, which could be rather small. In this way
the $\gamma$-ray photon flux reaching
the Earth is obtained by multiplying
$q_{\gamma}(r)$ by $\epsilon/4\pi r^2$ and integrating
the resulting quantity over the cloud volume along the line of sight.
Parameterizing a generic point P in the cloud volume by polar
coordinates
$(r, \theta, \phi)$, taking the Earth as the origin, it follows that
the
observed $\gamma$-ray flux per unit solid angle is
\begin{equation}
\Phi_{\gamma}(\theta,\phi)=\frac{\epsilon}
{4\pi} \int^{r_2}_{r_1} dr~ q_{\gamma}(r)
{}~, \label{eqno:51}
\end{equation}
where typical values of $r_1$ and $r_2$ are 10 kpc and 50 kpc,
respectively.
Assuming for the $H_2$ density distribution the same radial dependence
as
in Eq. (\ref{eqno:45}), Eq. (\ref{eqno:51}) becomes
\begin{equation}
\Phi_{\gamma}(\theta,\phi) =
\epsilon~ f~\rho_0~ I_1(\theta, \phi)~ I_2~, \label{eqno:52}
\end{equation}
where $f$ ($\sim 0.5$) denotes the fraction of dark matter in form of
$H_2$,
$\rho_0 \simeq 0.3~GeV~cm^{-3}$ is the local dark matter density and
\begin{equation}
I_1(\theta, \phi) = \int^{r_2}_{r_1} dr \left(\frac{a^2 + R_{GC}^2}
{a^2 + R^2} \right)^2~, \label{eqno:A5}
\end{equation}
\begin{equation}
I_2 = \int_{1~GeV}^{10^6~GeV} dE_p~\Phi_{CR}(E_p)~\sigma_{in}(p_{lab})
<n_{\gamma}(E_p)>~. \label{eqno:A6}
\end{equation}
The best chance, if any, to detect the $\gamma$-rays in question is
provided
by observations at high galactic latitude, and so we take
$\theta=90^0$.
The evaluation of $I_1(\theta,\phi)$ and $I_2$ is
straightforward and from Eq. (\ref{eqno:52}) we find
\begin{equation}
\Phi_{\gamma}(90^0) \simeq \epsilon~ 1.7 \times 10^{-6}~
{}~\frac{photons}{cm^2~ s~sr}~. \label{eqno:53}
\end{equation}
We notice that, although we have been
concerned with $\gamma$-ray production due to $pp$ collisions in
H$_2$ molecular clouds, the same effect occurs
for H atomic clouds as well. A possibility to discriminate
between the two components consists in performing a simultaneous search
for the 21-cm line, from which one can deduced the H abundance
(for more details see Danly et al. \cite{Danly}).

The inferred upper bound for $\gamma$-rays in the 0.8 - 6 GeV
range for high galactic latitude is $3 \times 10^{-7}$ $photons~
cm^{-2}~
s^{-1}~ sr^{-1}$ (Bouquet et al. \cite{Bouquet}).
Hence, we see from Eq. (\ref{eqno:53}) that the presence of
$H_2$ molecular clouds does not led at present to
any contradiction with the upper bound,
provided $\epsilon < 10^{-1}$.
An improvement of the sensitivity limit of next generations
of satellite-borne $\gamma$-ray detectors would yields
more stringent limits on $\epsilon$.\\

\section{Concluding remarks}
We have outlined a scenario for a baryonic dark halo,
in which the formation of MHO and/or $H_2$ molecular clouds in the
outermost
part of the halo naturally arises, because in a quiet ambient
the Jeans mass can reach values as low as $10^{-2}~ M_{\odot}$.
What is crucial
in discriminating the evolution of PGC clouds towards stellar globular
clusters or dark clusters are the decreasing collision rates and
UV-fluxes with increasing galactocentric distances. The most promising
way
to detect dark clusters of MHO is through correlation effects in
microlensing observations, as we expect the dark clusters not to have
been disrupted up to now.

A much more difficult task is the detection of $H_2$ molecular clouds.
A possible signature of such clouds would be their $\gamma$-ray flux
induced
by halo cosmic ray protons.
Our calculation gives only an upper bound which is not in conflict with
present detection limits.
A possible future detection might provide important informations
on the physics of the galactic halo (magnetic fields, cosmic ray flux,
molecular clouds). Another issue is the UV absorption lines
of extragalactic sources, a problem that
we have not addressed here and on which
we hope to come back in the future.\\

\acknowledgements{We would like to thank B. Bertotti, J. Lequeux,
F. Palla and D. Pfenniger
for useful discussions}

\end{document}